# Case report: Abnormal radiation dose rate measurement near Fukushima Daiichi nuclear power plant

Young Chan Seo [a*]

*[a] Department of Medical Device Development, Seoul National University College of Medicine, 103 Daehak-ro, Jongno-gu, Seoul, Republic of Korea*
*[*]Corresponding author: youngchanseo@snu.ac.kr*

## 1. Introduction

The Fukushima nuclear disaster occurred on Mar 11, 2011, at the Fukushima Daiichi Nuclear Power plant in Ōkuma, Fukushima, Japan. It was caused by a tsunami [1]. Since the severe accident, it has led to the release of high radioactive materials [2]. Radioactive leaks from the Fukushima nuclear accident have been in the air, soil, and sea. Environmental radiation measurements have been carried out in Japan, such as drone- and helicopter-based surveys, and vehicle-based and hand-carried measurements [3].

The author measured the air dose rate in several places, including near the Fukushima nuclear power plant site with the portable (hand-carried) survey meter, comparing with the reported values from the big institutes or national laboratories.

## 2. Methods and Results

This section describes the characteristics of the detector and the travel location (path). The value measured by the survey meter is compared with the known value to verify the value measured by the author.

This means that if a detector is valid for dosimetry in many situations, it is also valid for dose verification in a particular environmental situation.

### 2.1 Survey meter status

Table 1. Calibration Result, Measurement uncertainty is about 95% confidence level, k=2

| Source | Measurement Range | True Value ($\mu$Sv/h) | Measured Value ($\mu$Sv/h) | Calibration Coefficient | Averaged Calibration Coefficient | Measurement Uncertainty (%) |
|---|---|---|---|---|---|---|
| | ~ 100 $\mu$Sv/h | 30 | 29.7 | 1.011 | | |
| | | 50 | 49.6 | 1.008 | 1.013 | 7.4 |
| | | 70 | 68.7 | 1.019 | | |
| Cs-137 | ~1,000 $\mu$Sv/h | 300 | 293.4 | 1.022 | | |
| | | 500 | 505.7 | 0.989 | 1.009 | 7.4 |
| | | 700 | 689.7 | 1.015 | | |

The author measured the radiation dose rate on the air with the multi-Purpose survey meter (FH40-GL, Thermo Scientific, Massachusetts, United State). The survey meter was calibrated at the KORASOL, Seoul, South Korea on May 23, 2023. The Table 1. shows calibration status of the survey meter.

### 2.2 Travel Route

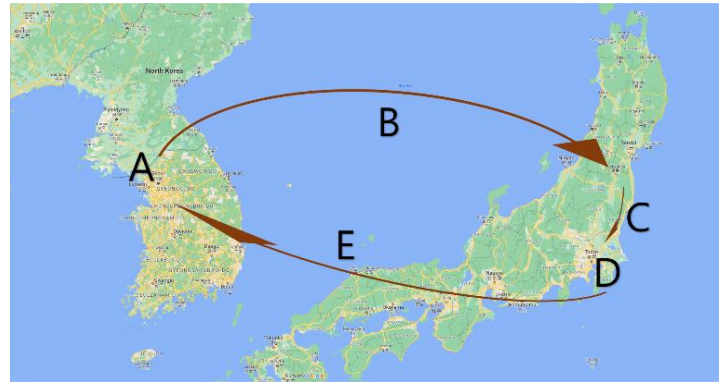

Fig. 1. Travel Route and travel spot (Google Map). (Spot A: Seoul), (Route B: Air route from Seoul to Sendai via the East Sea), (Spot C: Fukushima Prefecture, Japan), (Spot D: Tokyo, Japan), (Route E: Air route from Tokyo to Seoul via the East Sea).

Fig. 1 shows the travel route and spot where the author visited. It is described with the Google Map (https://maps.google.com/).

The author measured the radiation dose rates on July 14~23, 2023 (Spot A). On July 16, 2023, the author measured the radiation dose rate while flying from Seoul, Korea to Sendai, Japan via the East Sea (Route B). From July 17 to 20, 2023, the author measured radiation dose rates in Fukushima Prefecture, Japan (Spot C).

*2.3 Spot A, Seoul, South Korea.*

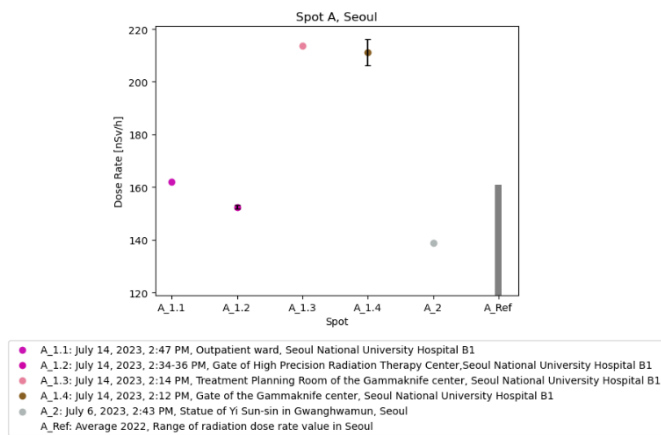

Fig. 2. Spot A, Seoul, South Korea. A single measurement was performed at Spot A_1.1, A_1.3, and A_2. Two repeated measurements were performed in Spot A_1.2 and A_1.4. A_Ref represents the range of the average annual spatial gamma dose rate measured at the 11 places in Seoul (2017 to 2022).

Both Spot A_1.1 and A_1.2 were measured in B1 floor, Seoul National University Hospital (서울대학교병원). The spots are in the range of A_Ref. Dose rates of Spot A_1.3 and A_1.4 were slightly higher than others though, and the values are still in the normal range of Korean environment [4] (No specific reference values were found in that Spot). A_2 was measured in Gwanghwamun (광화문), Seoul by author. A_Ref is range of the average annual spatial gamma dose rate measured at the 11 places in Seoul (Guro, Naegok, Nokbeon, Bongcheon, Samsung, Sanggye, Sinnae, Sincheon, Yongsan, Haengdang, Hwagok, 2017 to 2022) [5].

The measurement results in Spot A imply that the survey meter performed well for measuring radiation dose rate in Seoul, South Korea. This shows that the measurement of a random environment (air in Seoul) with the survey meter is valid. This indirectly shows that even if the dose rate measurements in other situations are abnormal, the survey meter is fine.

*2.4 Route B & E, Air route via the East Sea*

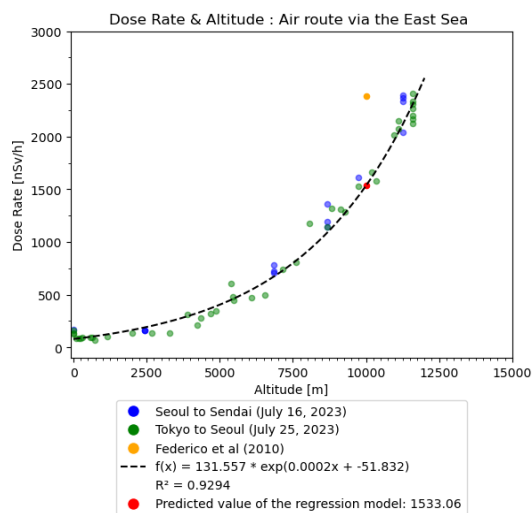

Fig. 3. Route B & E, Air route via the East Sea. The blue dots represent the measurements taken on the flight from Seoul, South Korea to Sendai, Japan (Route B). gThe green dots represent the measurements taken on the flight from Tokyo, Japan to Seoul, South Korea (Route E). Orang dot is the single value presented by Fedrico et al (2010). Red dot is predicted dose rate value by author, at 10,000m altitude.

The author took an OZ0152 Asiana flight from Seoul, South Korea to Sendai, Japan from 09:40 a.m. to 11:50 a.m. on July 16, 2023. In order to return to Seoul, South Korea from Tokyo, Japan, the author took an OZ101 Asiana Airline flight from 1:20 p.m. to 3:50 p.m. on July 25, 2023.

In Fig. 3, shows more details of the measurement. The measurements shown in the blue dots were taken 15 times only when the plane took off, climbed, and cruised (09:36 a.m. ~ 11:18 a.m.). The measurements shown in the green dots were taken 44 times by the plane from takeoff to landing (01:35 p.m. ~ 3:34 p.m.). Orange dot is presented by Fedrico et al and is a single value of the radiation dose rate at a commercial aviation of 10,000 m altitude. This value is calculated with the CARI-6 code for the Sao José dos Campos region of the SP in January 2008 [6]. Red dot is predicted single value by author, which is radiation dose rate at the 10,000 m altitude.

With measurements consisting of blue and green dots, the Exponential regression was calculated. The statistical code is written in python (version. 3.11.3) and SciPy library (version. 1.10.1) is used for statistical calculations.

The exponential regression model is as follows.

$$y = 131.557e^{0.0002x} - 51.832$$
$$(x: Altitude\ [m],\ y: Dose\ Rate\ [nSv/h])$$

The adjusted coefficient of determination ($R^2$) is 0.9294.

To compare with the values presented by Fedrico et al, the author calculated a single value according to the exponential regression model in the same altitude. It was calculated using the python's (version. 3.11.3) built-in Math module and SciPy library (version. 1.10.1).

The difference between the calculated value presented by the author and the calculated value by Fedrico et al. is approximately less than 35%, and the difference has existed, but the measurement value and model of the author are acceptable because the comparison target and the author did not assume a strict situation. Again, this shows that measuring a random environment (cosmic radiation) with the survey meter is valid and that there is no problem with the detector itself.

*2.5 Spot C Fukushima*

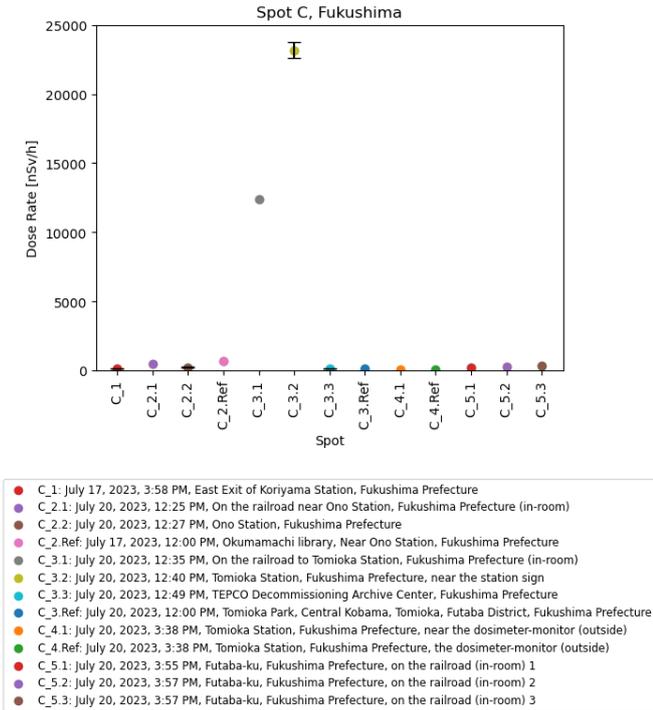

Fig. 3. Spot C Fukushima Prefecture, Japan.

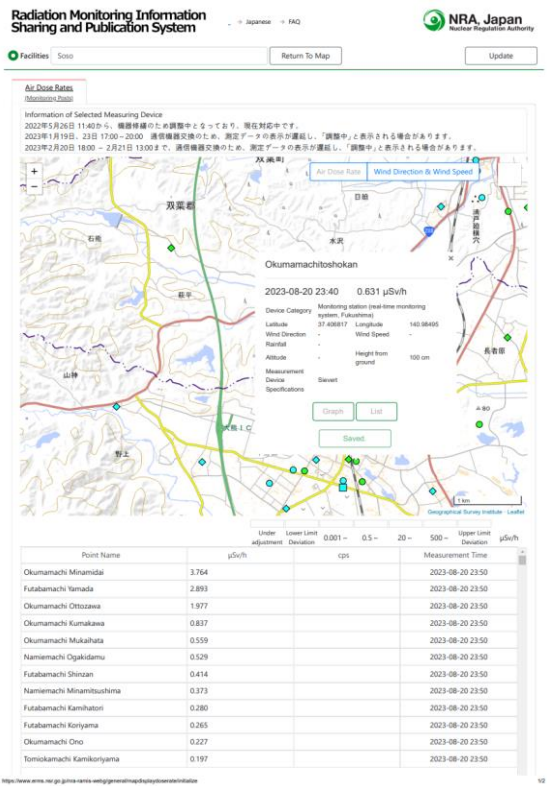

Fig. 4. At 11:40pm on 20 Aug 2023, Radiation dose rate at Okumamachitoshokan (Okumamachi Library, 大熊町図書館). Information provided by Radiation Monitoring Information Sharing and Publication System, Japan Nuclear Regulation Authority.

The author measured the radiation dose rate from July 17 to 20 with a survey meter at Fukushima Prefecture, and the measurements are shown in Fig. 3.

C_1 was measured three times repeatedly in Koriyama and no comparable reference values are presented.

C_2.1 (464 [$nSv/h$]) is a single value measured while traveling on a subway near the Ono Station (大野駅) and C_2.2 (248± 2 [$nSv/h$]) is three times repeated measurement on a stopped train at Ono Station.

C_2. Ref (711 [$nSv/h$] ) is the official data (Radiation Monitoring Information Sharing and Publication System, Japan Nuclear Regulation Authority) measured in the Okumamachi Library (大熊町図書館) with a difference of less than one hour from the time C_2.1 and C_2.2 were measured.

C_2. Ref data were obtained from the system shown in Figure 4. Since C_2.1 and C_2.2 measured by the author were measured in the train cabin, it is estimated that the values of C_2.1 and C_2.2 are lower than C_2. Ref, despite the similar location and time.

C_3.1 (12,359 [[$nSv/h$]) was measured once in the train moving to Tomioka Station (富岡駅), and C_3.2 (23,164 ± 586 [[$nSv/h$]) was measured 3 times at the Tomioka Station (outdoor). After moving by taxi from Tomioka station, the value of C_3.3 (129±1 [$nSv/h$]) was measured twice at the TEPCO Decommissioning Archive Center (東京電力廃炉資料館). C_3. Ref (163 [[$nSv/h$]) is obtained from the system as shown in Fig. 4. The C_3. Ref measurement was at the Tomioka Park (富岡公園).

C_4.1 (106 [[$nSv/h$]) is a single measurement and is in the same location as C_3.1, but the measurement time is different C_4. Ref (68 [[$nSv/h$]) is also a single measurement, the same location and time as C_4.1 , but with a difference of approximately 36%. C_4. Ref was acquired by the dosimeter-monitor installed at the Tomioka station.

On the train from Tomioka Station to Sendai Station, C_5.1, C_5.2 and C_5.3 were each single measured.

In Spot C, C_3.1 and C_3.2 measured around Tomioka station had unusually high dose rates, but there were no major problems in the other Spots.

## 2.6 *Spot D Tokyo, Japan*

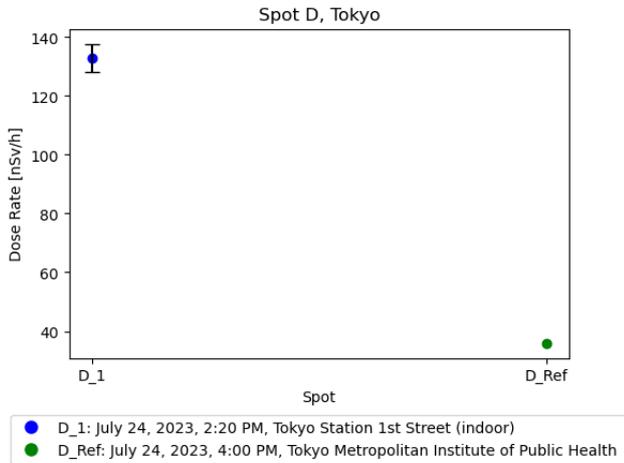

Fig. 5.  Tokyo, Japan.

Fig. 5 shows a big difference (under 75%) between D_1 and D_Ref. D_1 was measured 3 times, and the value of D_1 can be found in the system mentioned in Figure 4. Although there is a big difference, the location is very different. Since the dose rate of Tokyo Station (東京駅) could not be found, so the "Tokyo Metropolitan Institute of Public Health (D_Ref)" was mentioned. Even if it was the same location, it is a difference at a level that can be taken into account.

### 3. Conclusions

The author measured the dose rate with a Survey meter while traveling to Seoul (South Korea), the East Sea (flying), Fukushima (Japan), and Tokyo (Japan), and compared the measurements with existing references. The authors' measurements of random environments were valid, and the survey meter itself was not a problem.

An abnormal, very large dose rate was measured at Tomioka Station, about 10 km from the Fukushima Daiichi Power Plant.

Some Japanese institutions monitor the dose rate in many ways, but measurements are mainly in fixed places. Based on the case measurements by the author, it is recommended that Japanese institutions try to measure more carefully.

### REFERENCES

[1]    R. O. Gauntt *et al.*, "Fukushima Daiichi accident study: status as of April 2012," Sandia National Laboratories (SNL), Albuquerque, NM, and Livermore, CA …, 2012.

[2]    "Fukushima accident," in *Britannica*, T. E. o. Encyclopaedia, Ed., ed: Encyclopedia Britannica, 2023.

[3]    G. R. Jr, "Berkeley Lab's Advanced Monitoring Capabilities Still in Use 10 Years After Fukushima Earthquake and Nuclear Power Plant Disaster," in *NEWS FROM BERKERLY LAB*, ed, 2021.

[4]    "Korea Atomic Energy Research Institute's web page." https://www.kaeri.re.kr/env/radiationInfo/list?menuId=MENU00392 (accessed August, 2023).

[5]    "Environmental Radiation monitoring in Korea," KOREA INSTITUTE OF NUCLEAR SAFETY (KINS), KINS/ER-1183, December 2022, vol. 2. [Online]. Available: https://xedas.kins.re.kr/upload/report/04/er-028,vol.54.pdf

[6]    C. A. Federico, H. H. d. C. Pereira, M. A. Pereira, O. L. Gonçalez, and L. V. E. Caldas, "Estimates of cosmic radiation dose received by aircrew of DCTA's flight test special group," *Journal of Aerospace Technology and Management,* vol. 2, pp. 137-144, 2010.